\documentclass[%
 reprint,
 amsmath,amssymb,
 aps,prc,
]{revtex4-2}

\usepackage{graphicx}
\usepackage{dcolumn}
\usepackage{bm}
\usepackage{hyperref}
\usepackage{natbib}
\usepackage{amsmath}
\newcommand{\RNum}[1]{\uppercase\expandafter{\romannumeral #1\relax}}

\begin{document}


\title{An Indirect Measurement of $^6$Li(n,$\gamma$) Cross Sections}  

\author{Midhun C.V$^1$}\thanks{midhun.chemana@gmail.com}%
\author{M.M Musthafa$^1$}\thanks{mmm@uoc.ac.in}%
\author{S.V Suryanarayana$^{2,3}$}
\author{Gokuldas H$^1$}
\author{Shaima A$^1$}
\author{Hajara. K$^1$}
\author{Antony Joseph$^1$}
\author{T. Santhosh$^2$}
\author{A. Baishya$^2$}
\author{A Pal$^2$}
\author{P.C Rout$^2$}
\author{S Santra$^2$}
\author{P.T.M Shan$^1$}
\author{Satheesh B$^4$}
\author{B. V. John$^2$}
\author{K.C Jagadeesan$^5$} 
\author{S. Ganesan$^6$} 

\affiliation{$^1$ Dept. of Physics,University of Calicut, Calicut University P.O Kerala, 673635 India}
\affiliation{$^2$ Nuclear Physics Division, Bhabha Atomic Research Centre, Mumbai 400085, India}
\affiliation{$^3$ Manipal Centre for Natural Sciences, MAHE, Manipal - 576014, India}
\affiliation{$^4$ Dept. of Physics, Mahatma Gandhi Government Arts College, Mahe, 673311, India}
\affiliation{$^5$ Radiopharmaceuticals Division, Bhabha Atomic Research Centre, Mumbai 400085, India}
\affiliation{$^6$ Formarly Raja Ramanna Fellow, Bhabha Atomic Research Centre, Mumbai 400085, India}

\date{\today}

\begin{abstract}
The $^6$Li(n,$\gamma$)$^7$Li cross sections in the neutron energy range of 0.6 to 4 MeV have been measured by the experimental implementation of the direct capture formalism. This was done by measuring the $\gamma$ transition probability experimentally and accounting for the spin factor by theoretical calculation. The electromagnetic transition probabilities from $^7$Li$^*$ analogous to the initial neutron capture states of $^6$Li$+n$ were measured by populating the J$_i$ states of $^7$Li through $^7$Li($p,p'$)$^7$Li$^*$ reaction. The impact of coupling of resonant states, above neutron separation threshold of $^7$Li, in the neutron capture, is observed from the capture $\gamma$ spectrum. The measured cross sections were reproduced through {\sc fresco} and Talys-1.95 Direct Capture Calculations. 

\end{abstract}
\maketitle
Neutron induced reactions on Li isotopes are having a renewed interest due to their involvement in nuclear astrophysics and Gen.\RNum{4} nuclear reactors\citep{CPL6Ling,Li6Li71,Li67Li2}. Though $^6$Li(n,$\gamma$) reaction is having a significant importance, the reaction remains thus for unexplored. As the abundance ratio of $^6$Li to $^7$Li is used as an observable for estimating the time scale of stellar evolution, the reaction is having a major role in the Standard Big Bang Nucleosynthesis (SBBN) network calculations\citep{CPL6Ling,Malaney,Kenneth}. Further, $^6$Li(n,t) and $^6$Li(n,n$\prime$t) reaction contribute to the $^3$H breading in fusion reactors \citep{LibreadFusion}. In the emergency shut down system of fast neutron reactors, lithium enriched in $^6$Li is used. In these cases, $^6$Li(n,$\gamma$) reaction also will be initiated due to the high flux neutron environment and remains as the major source of uncertainty in the tritium production\citep{CrossUnSens}.
\paragraph*{} The abundances of $^6$Li and $^7$Li are highly influenced by the $^6$Li(n,$\gamma$) reaction rates. In order to avoid the wrong estimations and biasing in the SBBN network calculations, the cross section for this channel has to be accurately measured\citep{CPL6Ling,Malaney,Kenneth}. However, the $^6$Li(n,$\gamma$) reaction remains not well explored due to its small cross section as well as the lesser natural abundance of $^6$Li. The $^6$Li(n,$\gamma$)$^7$Li exhibits considerably small cross sections, in the order of $\mu b$, due to the predominance of $\alpha +t$ breakup mode, $^6$Li(n,$\alpha$)$^3$H channel being prominent. This breakup mode having a threshold of 2.47 MeV, dominates as the Q-value for the neutron capture of $^6$Li is as high as 7.25 MeV. Due to this higher Q-value, breakup levels populate more than the single particle levels. This suppresses the radiative neutron capture process in $^6$Li. However, the direct reaction component involving the single particle levels contributes significantly to the radiative neutron capture through Direct Capture (DC) mechanism \citep{dircap1,dircap2}.
\paragraph*{} There is only a single measurement on energy dependant cross section of $^6$Li(n,$\gamma$) existing in the literature, by Ohsaki et al., for the neutron energy range of 20 to 80 keV\citep{Ohsaki}. The other measurements available as exfor entries \citep{exfor}, by R. B Firestone et al. \citep{Firestone}, Chang Su Park et al. \citep{PARK} and G.A Batholomew et al. \citep{Bartholomew} are the Maxwellian averaged cross sections for thermal reactor neutron spectrum. Further, there are evaluations existing for $^6$Li(n,$\gamma$) in ENDF/B-\RNum{8}.0 \citep{ENDFB}, JEEF \citep{JEFF} and JENDL\citep{JENDL} basic evaluated nuclear data libraries. They are the derived data sets estimated by fitting the R-Matrix formalism on $^6$Li(n,t)$^4$He cross sections\citep{ENDFB}. Due to the limitations with the $^6$L target and lower cross sections, the direct measurement of $^6$Li(n,$\gamma$) is potentially challenging.
\paragraph*{}As a solution to this problem, an experimental methodology has been implemented by utilizing the properties of the Direct Capture formalism. The measurement of $^6$Li(n,$\gamma$) cross sections has been attempted through this method, by populating the initial neutron capture states by inelastic scattering of protons on $^7$Li. The $\gamma$ transitions from the $^7$Li initial state to bound states of $^7$Li are measured by generating $p-\gamma$ coincidences. The capture cross sections were determined through the approach of Direct Capture mechanism, reformulated for the experimental implementation. The current experiment is having resemblances with the established Oslo method\citep{oslo1,oslo2}. In this study, only a single large volume LaBr$_3$ scintillator was used instead of $\gamma$ calorimeter. This measurement has been successful because all the $\gamma$ events from the initial neutron capture states are well resolvable and $\gamma$ multiplicity is near to unity. 


\paragraph*{}The direct capture model (DC) is well established for explaining the radiative captures at lower energies, and neutron capture reactions for light elements, where the compound nuclear level density models are insufficient to explain the radiative capture\citep{dircap2}. The DC accounts for the capture cross section as the sum of direct capture to the discrete states and the single particle continuum states. The Direct Capture formalism is expressed in its own form based on Ref. \citep{dircap1} and Ref. \citep{dircap2},
\begin{equation}
\small
\sigma^{DC}(E) = \sum_{f=0}^{f_{max}}S_{f}\sigma^{dis}_f(E) + \left\langle S \right\rangle \int_{E_{0}}^{E_{n}} \rho(E_f,J_f,\pi_f)\sigma^{con}_f(E_f)dE_f
\end{equation}
The first part in Eq. 1, $\sum_{f=0}^{f_{max}}S_{f}\sigma^{dis}_f(E_f)$ corresponds to the direct capture to the discrete resonant states defined as $f$. $S_{f}$ is the spectroscopic amplitude corresponding to each resonant state. Second part, $\left\langle S \right\rangle \int_{E_{0}}^{E_{n}} \rho(E_f,J_f,\pi_f)\sigma^{con}_f(E_f)dE_f$ corresponds the capture to the single particle continuum states. The $\left\langle S \right\rangle$ is the average spectroscopic amplitude of the continuum states distributed from the continuum formation threshold $E_{0}$ to the neutron separation energy $E_{n}$. $\rho(E_f,J_f,\pi_f)$ is the single particle level densities corresponding to each $E_f$, which are also a function of the spin and parity of the state. The $\sigma^{con}_f(E_f)$ stands for the capture corresponding to the each of $f$ states in the continuum having energy $E$, which is assumed to be discretized in an interval $dE_f$. The $\sigma^{con}_f(E_f)$ can be expressed, based on the electric and magnetic modes of transition, as
\begin{equation}
\small
\begin{split}
\sigma^{con}_f(E_f) = &\frac{2J_f+1}{Ek(2J_A+1)(2J_n+1)}  \\
\times & \sum_{I_f,J_i,l_i,I_i}\left[ C_1 k_{\gamma}^3 (|M_{E1}|^2+|M_{M1}|^2)+  C_2 k_{\gamma}^5|M_{E2}|^2 \right]
\end{split}
\end{equation}
Here $J_f$ is the spin of the final state after capture followed by the electromagnetic transition, $J_A$ is the spin of the target nuclei and $J_n$ is the spin of neutron. $E$ and $k$ are respectively energy and wave number of neutrons corresponding to $J_i$. $C_1$ and $C_2$ are the  normalization constants corresponding to dipole and quadrupole transitions. $M_{E1}$, $M_{M1}$ are the electric and magnetic dipole matrix elements and $M_{E2}$ is the matrix element corresponding to electric quadrupole transition. $k_{\gamma}$ stands for the wave number of the emitted photon. 

\begin{figure}
\includegraphics[width=\columnwidth]{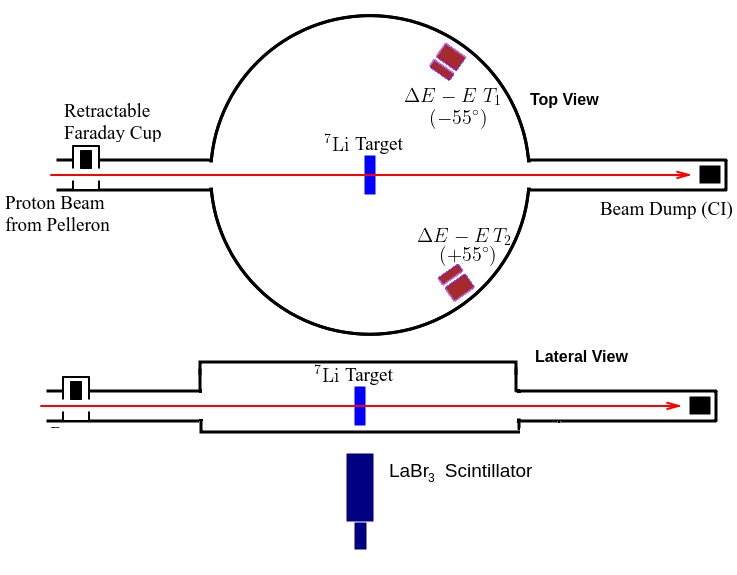}
\caption{Illustration of experimental setup in top and lateral views}
\end{figure}

\begin{figure}
\includegraphics[width=\columnwidth]{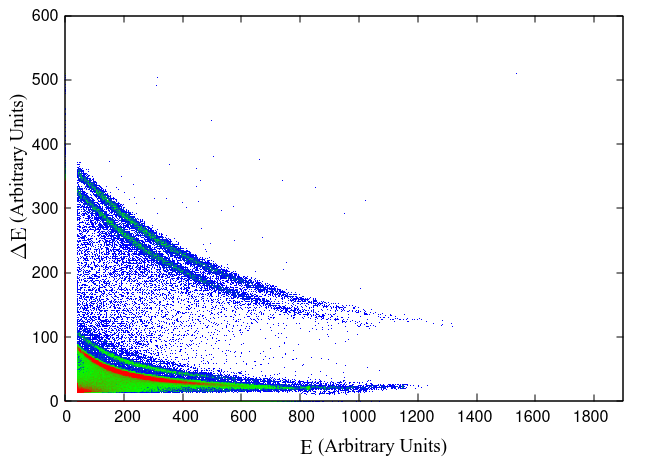}
\caption{The typical $E-\Delta E$ Correlation plot recorded in the telescope}
\end{figure}

\paragraph*{}The neutron capture in $^6$Li populates initial capture state with an excitation energy of 7.25 MeV + E$_n (\frac{A}{A+1})$. This energy is well above the $^7$Li $\rightarrow \alpha + t$ breakup threshold of 2.47 MeV. However, the transition from $J_i$ to the $\alpha + t$ breakup continuum levels will not contribute to the radiative capture process. The population of breakup states results to the continuum breakup of $^7$Li.  Hence, the $3/2^-$ and $1/2^-$ states corresponding to ground and 477 keV are the only $J_f$ states involving the radiative capture. The present formulation assumes that the $3/2^-$ and $1/2^-$ states are also a part of the single particle continuum states. However, for these states, the $\rho(E_f,J_f,\pi_f)$ term generates the definite shape of the resonance. This assumption is taken for modifying Eqn. 1 to a single integral ranging from 0 to E$_{n}$. Hence the reformulated form can be expressed as, 
\begin{equation}
\small
\begin{split}
\sigma^{DC}(E) = \frac{ (2J_f+1)}{Ek(2J_A+1)(2J_n+1)} \int_{0}^{E_{n}} \rho(E_f,J_f,\pi_f) & \\  \times \sum  C_1 k_{\gamma}^3  (|M_{E1}|^2+|M_{M1}|^2)+C_2 k_{\gamma}^5|M_{E2}|^2 dE_f
\end{split}
\end{equation}
In the following part of the article, $\frac{ (2J_f+1)}{Ek(2J_A+1)(2J_n+1)}$ is expressed as $g_j$, the spin factor,  and $\int_{E_{x}}^{E_{max}} \rho(E_f,J_f,\pi_f)  \\  \times \sum  C_1 k_{\gamma}^3  (|M_{E1}|^2+|M_{M1}|^2)+C_2 k_{\gamma}^5|M_{E2}|^2 dE_f$ as $\Gamma (E_x)$, gamma strength term.
\paragraph*{} The gamma rays produced from each excitation energy, equivalent to the excitation energy of the neutron capture, having a capture $\gamma$ distribution depend on the transitions to $J_f$ states. The integral of this distribution is considered as $\Gamma (E_x)$, which has been measured in the current experiment. The events populating resonant and direct breakup levels, during the collective excitation, are not producing any gamma rays. Thus the $\gamma$ spectra produced in inelastic excitation of $^7$Li are purely from the transition of the $J_i$ states to the bound states. Thus in the current study, $\Gamma (E_x)$ is measured for the equivalent neutron energies using inelastic scattering of protons on Li target. The $g_i$ has been calculated theoretically for obtaining the cross sections.


\begin{figure}
\includegraphics[width=\columnwidth]{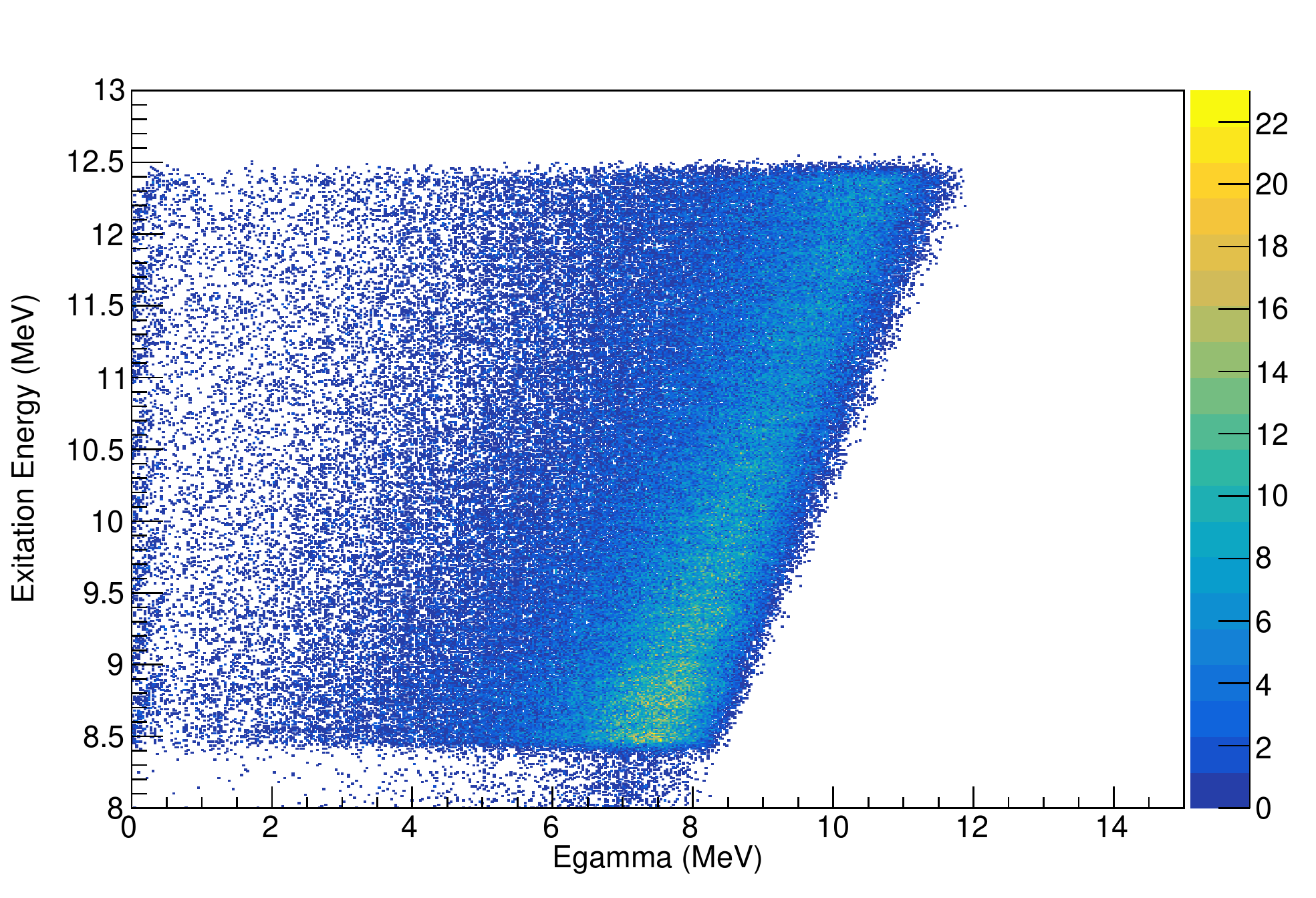}
\includegraphics[width=\columnwidth]{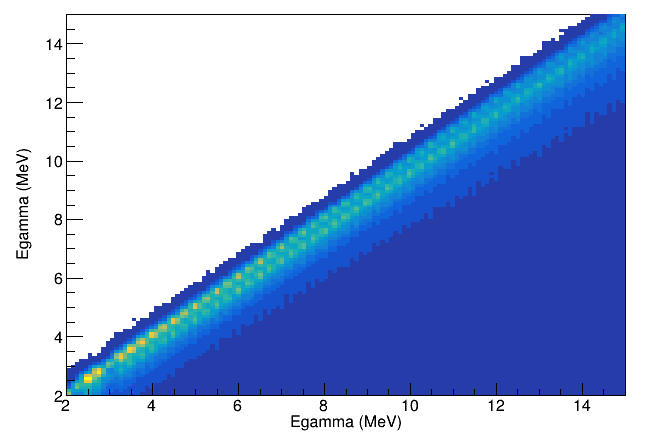}
\includegraphics[width=\columnwidth]{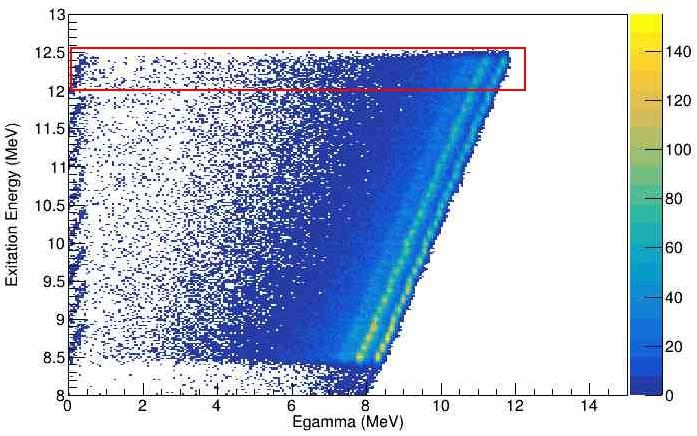}

\caption{$(a)$: The original E$_\gamma$-E$_x$ correlation, $(b)$ : The response matrix for LaBr$_3$ scintillation detector, $(c)$: The unfolded E$_\gamma$-E$_x$ correlation matrix with the gate defined for 12.375 MeV excitation
}

\end{figure}

\paragraph*{}The experiment is performed using 16 MeV proton beam from BARC-TIFR Pelletron facility, Mumbai, India. A $^{nat}$Li target of 1.4 mg/cm$^2$ was used for the experiment. The experimental setup has been illustrated in Fig 1. Two silicon detector telescopes having 25$\mu$m and 1500$\mu$m $\Delta$E-E pairs, were mounted at $+55^{\circ}$ $-55^{\circ}$ at a distance of 7.5cm from the target to record the inelastically scattered protons. A 3$^\prime$ diameter and 7$^\prime$ length LaBr$_3$ scintillator, coupled to a fast photomultiplier tube, was placed at 90$^\circ$, at the bottom side of the target. This was used for measuring the $\gamma$ spectra. The distance between the target and LaBr$_3$ detector was 12 cm.

\paragraph*{}The inelastically scattered protons were identified from the E-$\Delta$E correlation plot by proper gating. The $\gamma$ events in the LaBr$_3$ scintillator, with the elastically scattered protons are identified. The events were converted to the excitation energy of $^7$Li from the energy of elastically scattered protons, using two body kinematics. A correlation matrix between $\gamma$ energy (E$_\gamma$) and excitation energy of $^7$Li (E$_x$) was constructed. This matrix was unfolded to original $\gamma$ matrix using response matrix for the current LaBr$_3$ detector. The unfolding procedure was done using in-house developed machine learning assisted unfolding algorithm by preserving the excitation energy information of the events\citep{unfold}. The obtained E-$\Delta$E correlation plot is presented in Figure. 2. The measured correlation, simulated response matrix for the detector and unfolded $E_\gamma$-$E_x$ matrix are shown in Figure. 3.

\paragraph*{}The unfolded $\gamma$ matrix has  been appropriately gated for the excitation energy and projected to the $\gamma$ spectrum. The spectrum holds three colonies of gammas, evident in all excitation energies. Discrete $\gamma$ colonies are corresponding to the electromagnetic transitions from J$_i$ state with energy $E_x$ to the $ 3/2^-(g.s),1/2^-(477keV)$ states below $\alpha + t$ breakup threshold. The continuum $\gamma$ colony is identified as emerging due to the coupling of $ 7/2^-, 5/2^-$ and $3/2^- $ resonant levels, above neutron separation threshold of $^7$Li, to the electromagnetic transitions to the $3/2^-(g.s)$ or $1/2^-(477keV)$ states. These states are also overlapped by the $\alpha + t$ breakup continuum, results a higher decay width. The inelastic scattering populates the collective levels than the single particle levels in nuclei. In the case of $^7$Li, there are $\alpha + t$ breakup continuum present in the region of the collective levels. The population of the breakup levels seldom contributes to the $\gamma$ production, as they end up with the breakup (the inelastic breakups). Hence the $\gamma$ events observed in coincidence with the inelastically scattered protons can be assured as originated from the single particle levels.

\begin{figure}
\includegraphics[width=\columnwidth]{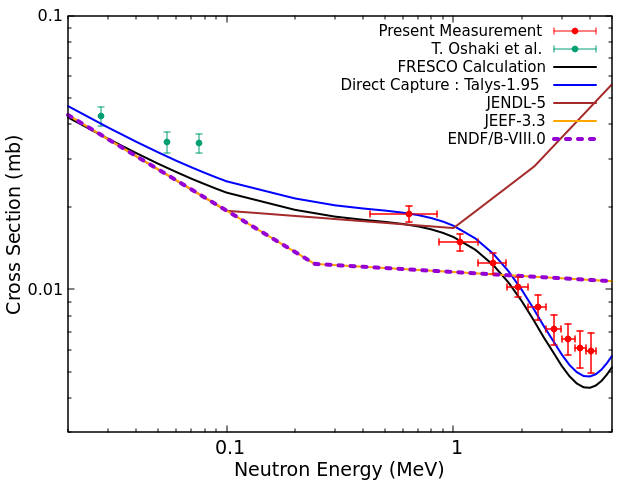}
\caption{Measured $^6$Li$(n,\gamma$) cross sections compared with Direct Capture Model Calculations in {\sc fresco}, Talys-1.95, and the evaluations by ENDF/B-VIII.0, JEEF-3.3 and JENDL-5 libraries }
\end{figure}

\begin{figure}
\includegraphics[width=\columnwidth]{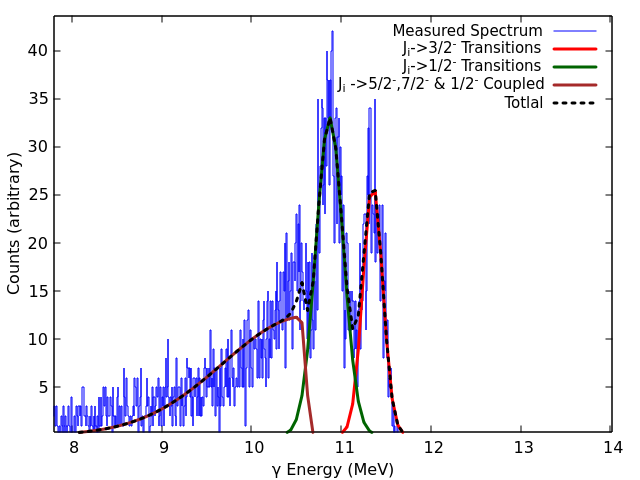}
\caption{Measured $\gamma$ spectrum at 11.5 MeV excitation energy along with the calculated spectrum}
\end{figure}

\paragraph*{}The total counts in the $\gamma$ colonies are retrieved and normalized with the inelastically scattered protons corresponding to the each excitation energy bin. The normalized $\gamma$ yield is corrected for the solid angle between the LaBr$_3$ scintillator and the target. The attenuation in the bottom lid of the scattering chamber is also accounted based on Geant4 simulation \citep{Geant4}. The intrinsic efficiency of the detector has already been applied through the response matrix during the unfolding process. This corrected normalized yield is the measure of $\Gamma (E_x)$ in the Direct Capture Formalism. The spin factor, $g_i$, is calculated using the direct capture calculation module incorporated in Talys-1.95. This provides an accurate estimation of the wave number, k. Optical potential for n+$^6$Li is taken from S. Chiba et al \citep{Chiba} for the calculation of the spin factor. The $^6Li(n,\gamma)$ cross sections were then deduced as $g_i \Gamma (E_\gamma)$.

\paragraph*{}Theoretical calculations are performed to reproduce the direct capture cross sections as well as to validate the measured $\gamma$ spectrum. A coupled channel method has been adopted for the calculation. $n+^6$Li is taken as the entrance channel mass partition and $^7$Li$+\gamma$ partition in the exit channel. The $n+^6$Li potentials by Chiba et al.\citep{Chiba} was used for the calculations. The calculations were performed using {\sc fresco} coupled channel code \citep{fresco}. The problem is defined as to calculate the direct radiative transitions from $J_i$ to $3/2^- \: (g)$ and $1/2^- \: (477\;keV)$ levels. The levels above neutron separation energy is considered as coupled to the $3/2^-$ and $1/2^-$ levels. Further, the overlap of $\alpha + t$ on the resonant states above the breakup threshold has been considered. The breakup continuum is discretized with $\alpha + t$ relative energy bin of 0.5 MeV, and the spin is calculated as the sum of spin of $t$ and the relative angular momentum of the bin. Theoretical spectroscopic factor for $^6$Li+n from the shell model calculations by Cohen and Kurath have been used for the resonant states\citep{COHEN}. An optimized average spectroscopic factor of 0.75 was used for $\alpha + t$. The {\sc fresco} calculation has been limited to E$_1$ and M$_1$ mode of $\gamma$ transitions and higher multipolarities have been neglected. A separate Monte-Carlo calculation, based on the {\sc fresco} calculated cross sections, has been performed to reproduce the experimentally measured $\gamma$ spectra.
\paragraph*{}Along with the {\sc fresco} calculations, statistical Direct Capture calculations were performed using Talys-1.95 nuclear reaction code \citep{talys}. This is to account for the compound nuclear, pre-equilibrium and the continuum component of the direct capture mechanism. The Microscopic level densities (temperature dependent HFB, Gogny force) from Hilaire’s combinatorial tables were used for the calculations of compound nuclear contribution as well as the continuum component of direct capture\citep{HilaireLD}. The predictions by Talys calculations were also compared with experimental results.


\paragraph*{}The measured cross sections for $^6$Li(n,$\gamma$) are illustrated as locally averaged histogram values in Figure 4, along with the measurement by T. Oshaki et al., for comparison. Theoretical calculations using {\sc fresco} and Talys-1.95 are also presented in Figure 4. Present measurement of cross sections for $^6$Li(n,$\gamma$) is limited to the Direct Capture, and the compound nuclear contributions. Due to the limitations in the methodology, preequilibrium contribution is not addressed in the experiment. Measured cross sections are in the $\mu$b range and cover, for the first time, energies of 0.6 to 4 MeV as compared to measurements of Oshaki et al. which are in the range of 30 to 70 keV energy region.

\paragraph*{} The {\sc fresco} calculated discrete component of the Direct Capture is illustrated as the black solid line in Figure. 4. The calculations reproduce the experimental cross sections within the error bars, for the spectroscopic factors by Kohan and Kurath. This indicates that the major contribution of the $^6$Li(n,$\gamma$) is from the discrete component of the direct neutron capture. Further, there exists a strong coupling of $5/2^-$, $7/2^-$ and $3/2^-$ levels, which lies above the neutron separation threshold of $^7$Li, on $J_i \rightarrow 3/2^-$ and J$_i \rightarrow 1/2^-$ transitions. These resonant states are overlapping with the $\alpha+t$ breakup continuum and indicates a larger energy width. This coupling effect is emerged as the low energy tail along with $\gamma$ colonies, corresponding to $J_i \rightarrow 3/2^-$ and J$_i \rightarrow 1/2^-$ transitions, in the experimental $\gamma$ spectrum. The experimental $\gamma$ spectrum has been well reproduced using the Monte-Carlo simulation based on the {\sc fresco} calculation and is presented in Figure 5. This assures the validity of couplings considered for the calculation of discrete components of the direct capture in $^6$Li(n,$\gamma$) reaction.

\paragraph*{}Talys-1.95 calculated continuum component of Direct Capture, Houser-Feshback + Pre-equilibrium components are much lesser compared to the discrete component of the direct capture. Talys calculated discrete component of Direct Capture in $^6$Li(n,$\gamma$) is also well in agreement with the experimental cross sections and {\sc fresco} calculations. The inhibition of the compound nuclear component in $^6$Li(n,$\gamma$) reaction is due to the lesser number of levels available in the $^7$Li compound nucleus. Further the strong coupling of $\alpha + t$ breakup is also reducing the capture cross section significantly. The contributions of E1, M1 and E2 mode of $\gamma$ transitions to the total direct capture are presented in Figure 7. Compared to the E1 mode of transition, the M1 and E2 transition yields are much lower and not contributing significantly to the total radiative capture. The sum of Direct Capture, Houser-Feshback and Pre-equilibrium is well reproducing the experimental cross sections for $^6$Li(n,$\gamma)$. The Direct Capture, Houser-Feshback and Pre-equilibrium components of neutron capture along with experimental cross sections are illustrated in Figure. 6.

\paragraph*{} Excitation functions available with ENDF/B-\RNum{8}.0, JEEF-3.2 and JENDL-5 are compared with the experimental cross sections and theoretical predictions. The comparison is presented in Figure 4. ENDF/B-VIII.0 and JEEF-3.2 cross sections are identical and not reproducing the theoretical or experimental cross sections above 0.1 MeV. At lower energies, below 0.1 MeV, all the evaluations are matching with the discrete component of the direct capture. But at energies above 0.1 MeV, there is a huge discrepancy between ENDF/B-VIII.0,JEEF-3.2 and JENDL-5. The JENDL-5 is highly over predicts the experimental data.  Further, the shape of the excitation function differ significantly from the measured and calculated curves. $^6$Li(n,$\gamma$) evaluated data in ENDF/B are obtained by fitting the R-Matrix on the $^6$Li(n,$\alpha$), the breakup channel. Hence it is anticipated as this fitting may be inappropriate for accurate parametrisation of the direct capture.

\begin{figure}
\includegraphics[width=\columnwidth]{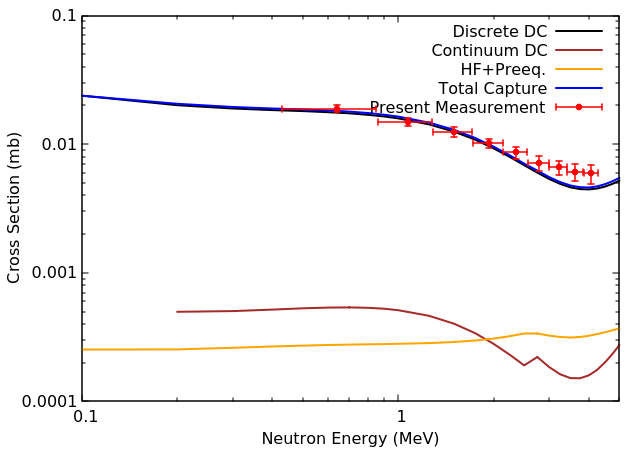}
\caption{Discrete and Continuum components of direct capture and the Hauser-Feshbach- + pre-equilibrium component (HF+Preeq) with the the total cross section compared with the experimental data}
\end{figure}
\begin{figure}
\includegraphics[width=\columnwidth]{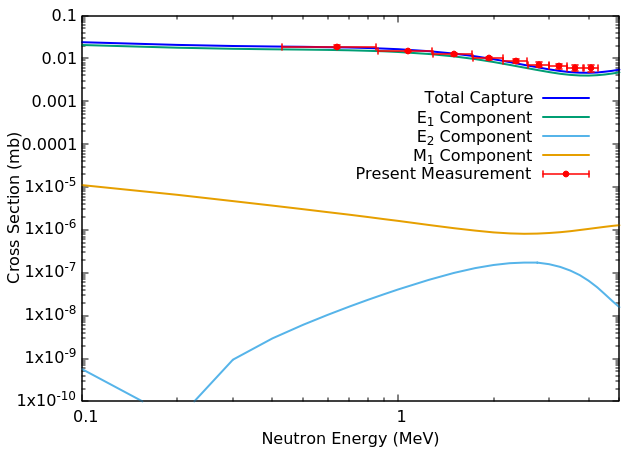}
\caption{Contribution of E$_1$, E$_2$ and M$_1$ modes of electromagnetic transition to the total capture cross section.}
\end{figure}
\paragraph*{}Summarizing, the $^6$Li(n,$\gamma$) cross sections ave been deduced by measuring the $\gamma$ transition probability experimentally and by accounting for the spin factor by theoretical calculation. The initial neutron capture states have been populated through the inelastic scattering of protons in $^7$Li. The method is analogous to the established Oslo method used to measure the neutron capture. However, this current measurement is fulfilled with a single LaBr$_3$ scintillator, than the total absorption spectrometers ($\gamma$ calorimeters). This is because of the average multiplicity of the $\gamma$ events from $J_i$ states covered in this study, are near to unity and the primary $\gamma$ events were well resolved. The accessibility of this method to the direct capture mechanism is assured by comparing the measured $\gamma$ spectrum to the calculated $\gamma$ spectrum. This shows that the major mechanism involved in $^6$Li(n,$\gamma$) is the discrete component of Direct Capture with E1 mode of transition. Continuum component of Direct Capture and HF+Prequilibrium are significantly smaller compared to the discrete component. E2 and M1 components a re significantly lesser compared to the E1 mode of transition. The calculations indicate there exist a significant coupling of the $5/2^-, 7/2^-$ and $3/2^-$ levels, above neutron separation threshold of $^7$Li, to the $J_i\rightarrow 3/2^-$ and $J_i\rightarrow 1/2^-$ transitions in $^6$Li(n,$\gamma$) reaction.
\section*{Acknowledgements}
The support of BARC-TIFR Pelletron-LINAC group during the experiment is acknowledged.The authors acknowledge the staff in TIFR Target Laboratory in the development of Li target for the experiment. The authors acknowledge A. Shanbhag, Health Physics Division, BARC for their kind support during the experiment. This study is part of the project supported by DAE-BRNS, Sanction Order 36(6)/14/30/2017-BRNS/36204.

\bibliography{Refer}
\end{document}